\documentclass[preprintnumbers,amsmath,amssymb,floatfix,10pt,prd,twocolumn,
superscriptaddress,nofootinbib]{revtex4-2}
\usepackage{latexsym}
\usepackage{epsfig}
\usepackage{epstopdf}
\usepackage{graphicx}
\usepackage{amssymb}
\usepackage{amsmath}
\usepackage{etoolbox}
\AfterEndEnvironment{strip}{\leavevmode}
\usepackage{amsfonts}
\usepackage{subfigure}
\usepackage{dcolumn}
\usepackage{bm}
\usepackage[normalem]{ulem}
\usepackage[dvipsnames]{xcolor}
\usepackage{hyperref}
\usepackage{color}
\usepackage{comment}
\usepackage{float}
\usepackage[utf8]{inputenc}
\hypersetup{colorlinks=true,linkcolor=blue,citecolor=red,urlcolor=magenta}
\usepackage{mathrsfs}
\usepackage{mathtools}
\usepackage{orcidlink}
\usepackage{soul}

\begin{document}

\title{Non-Schwarzschild black holes sourced by scalar-vector fields}

\author{Manuel Gonzalez-Espinoza \orcidlink{0000-0003-0961-8029}}
\email{manuel.gonzalez@pucv.cl}
\affiliation{Instituto de F\'{\i}sica, Pontificia Universidad Cat\'olica de 
Valpara\'{\i}so, 
Casilla 4950, Valpara\'{\i}so, Chile}

\author{Y. Gómez-Leyton}
\email{ygomez@ucn.cl}
\affiliation{Departamento de Física, Universidad Católica del Norte, Av. Angamos 0610, Antofagasta, Chile.}

\author{Z. Stuchlik}
\email{zdenek.stuchlik@fpf.slu.cz}
\affiliation{Research Centre for Theoretical Physics and Astrophysics, Institute of Physics, Silesian University in Opava,
Bezrucovo nam. 13, 74601 Opava, CZ 
}

\author{Francisco Tello-Ortiz \orcidlink{0000-0002-7104-5746}}
\email{francisco.tello@ufrontera.cl}
\affiliation{Departamento de Ciencias Físicas, Universidad de La Frontera, Casilla 54-D, 4811186 Temuco, Chile.}

\begin{abstract}
{In this work, we construct a static and spherically symmetric black hole solution sourced by a nonlinear scalar-vector sector within the framework of the Minimal Geometric Deformation (MGD) approach. Starting from a scalar-vector gravity theory in which the vector field is described by nonlinear electrodynamics, the gravitational field equations are integrated to obtain a non-Schwarzschild geometry. We show that the resulting family of solutions naturally splits into two black hole branches with distinct physical properties. While one branch satisfies the odd- and even-parity stability criteria, it violates the weak energy condition outside the event horizon. Conversely, the complementary branch exhibits a more satisfactory effective matter sector, admitting a real scalar-field reconstruction and a regular nonlinear electromagnetic coupling, although it does not satisfy the odd-parity stability criterion. The causal structure and stability properties of both black hole branches are analyzed and compared, whereas the matter reconstruction, geodesic motion and thermodynamic properties are investigated for the branch exhibiting the most satisfactory effective matter sector. These results reveal the rich structure of the MGD solution space and illustrate the interplay between dynamical stability, energy conditions and causal structure.}
\end{abstract}

\maketitle

\section{Introduction}

The theoretical and observational study of black holes (BHs) continues to be a cornerstone of 
modern gravitational physics. Over the past decade, the field has experienced an unprecedented 
renaissance driven by two major developments: the direct detection of gravitational waves from 
compact binary coalescence by LIGO-Virgo-KAGRA 
\cite{LIGOScientific:2016aoc,LIGOScientific:2021djp}, and the imaging of supermassive BH
shadows by the Event Horizon Telescope (EHT) 
\cite{Akiyama:2019cqa,Akiyama:2019eap,Akiyama:2022tqa,Vagnozzi:2022moj}. 
These advances have opened new avenues for testing the strong-field predictions of general 
relativity (GR) and for probing possible signatures of additional fields or new gravitational 
degrees of freedom.

From a theoretical perspective, there is a longstanding interest in black hole solutions supported 
by scalar, vector, or non-linear electrodynamic fields. Such matter sectors arise naturally in 
effective field theories beyond GR, including low-energy string-inspired models 
\cite{Gibbons:1987ps,Garfinkle:1990qj}, generalized Proca theories 
\cite{Heisenberg:2014rta,deFelice:2017paw}, non-linear electrodynamics 
\cite{Hendi:2012um,Bronnikov:2000vy}, and scalar-vector-tensor extensions 
\cite{Kase:2018aps,Babichev:2023psy}. In this context,
BH solutions supported by anisotropic matter and non-linear
electromagnetic sectors have been actively investigated, showing that
non-standard energy-momentum configurations can consistently modify
the geometry and stability properties of the spacetime
\cite{Kim:2025sdj,Lee:2026cit}. These constructions provide a rich framework to explore 
phenomena such as modified horizon geometries, deviations in geodesic motion, new stability channels, and non-standard thermodynamics. However, obtaining exact and physically well-behaved solutions, particularly those preserving asymptotic flatness and a single horizon, remains a 
highly non-trivial task.

The gravitational decoupling (GD) program, and specifically its Minimal Geometric Deformation (MGD) formulation introduced in \cite{Ovalle:2017fgl}, offers an elegant and systematic 
approach for constructing new solutions to the Einstein equations by deforming a known ``seed'' geometry. In this framework, the total energy--momentum tensor is decomposed into a seed sector and 
an auxiliary ``decoupling'' contribution, with the deformation encoded directly in the metric. 
A key feature of the MGD method is the possibility of implementing \emph{minimal} deformations, where the temporal metric component is preserved, and only the radial part receives a correction. 
This ensures that many geometric properties of the seed solution, including its causal structure 
and asymptotic behavior, are maintained while still allowing for non-trivial modifications 
induced by additional matter sources. This approach has been widely applied in the context of BHs physics
\cite{Ovalle:2018ans,Contreras:2019iwm,daRocha:2019pla,Heydarzade:2023dof,Ovalle:2018umz,Cavalcanti:2022adb,Cavalcanti:2022cga,Meert:2021khi,Sultana:2021cvq,Ovalle:2020kpd,daRocha:2020gee,Fernandes-Silva:2019fez,Casadio:2022ndh,Ovalle:2022eqb,Estrada:2021kuj,Ovalle:2023ref,Zhang:2022niv,Khosravipoor:2023jsl,Contreras:2021yxe,Ramos:2021jta,Ovalle:2021jzf,Arias:2022jax,Avalos:2023ywb,Avalos:2023jeh,Ovalle:2023vvu,Casadio:2023iqt,Contreras:2018nfg,Estrada:2020ptc,Fernandes-Silva:2019fez,daRocha:2020gee,Estrada:2021kuj}, demonstrating its versatility.

BHs impose stricter geometric and physical requirements, such as 
the preservation of a consistent Lorentzian signature, the existence of a unique 
event horizon, and compatibility with perturbative stability conditions. These constraints make the development of minimally deformed BHs particularly appealing: such solutions maintain 
the essential structure of a classical BH while incorporating the influence of additional fields in 
a controlled and analytically tractable manner. Moreover, minimally deformed BHs offer a 
clean theoretical setting for examining how new matter sectors modify observable quantities such as 
orbital precession, lensing, quasinormal spectra, or thermodynamic relations, without altering the 
overall causal framework.

In the present work, we construct a family of static and spherically symmetric BH solutions sourced by 
a non-linear scalar-vector sector and generated through the MGD approach. The resulting geometry 
preserves the temporal metric component of the Schwarzschild solution and introduces a 
well-defined deformation exclusively in the radial sector. To our knowledge, this represents one of 
the first explicit instances in which a minimally deformed extension of Schwarzschild leads 
unambiguously to a single-horizon, asymptotically flat, and physically interpretable BH
supported by a non-linear scalar--vector interaction. This provides a new analytical laboratory to 
explore the interplay between geometric deformations, non-linear matter fields, and classical 
BH phenomenology, and contributes to the broader effort of extending GR solutions using the 
GD methodology.

The article is organized as follows: Sect. \ref{section2} revisited GD by MGD in short. Sect. \ref{section3} introduces the model, its causal structure, reconstruction of the matter sector, and stability analysis via odd and even parity master constraints. Then, Sect. \ref{section4} presents the geodesic motion study for massive particles and Sect. \ref{section5} is devoted to the thermodynamic study. Finally, Sect. \ref{section6} concludes the work.

Throughout this work we adopt metric signature 
$\{- , + , + , +\}$ and geometric units $c = G = 1$. We follow the sign 
conventions used in \cite{Misner:1973prb}, with the Riemann tensor defined as
$
R^{\rho}{}_{\sigma\mu\nu} = \partial_{\mu}\Gamma^{\rho}_{\nu\sigma}
- \partial_{\nu}\Gamma^{\rho}_{\mu\sigma}
+ \Gamma^{\rho}_{\mu\lambda}\Gamma^{\lambda}_{\nu\sigma}
- \Gamma^{\rho}_{\nu\lambda}\Gamma^{\lambda}_{\mu\sigma},
$ and
the Ricci tensor by $R_{\mu\nu} = R^{\rho}{}_{\mu\rho\nu}$.

\section{Gravitational decoupling in short}\label{section2}

Here we introduce the main aspect to be used in the context of the well-known GD by MGD approach \cite{Ovalle:2017fgl,Ovalle:2017wqi,Ovalle:2018ans}. The main point is to consider an extended energy-momentum tensor, composed by two sources. Explicitly this reads
\begin{align}\label{StressTensorEffective}
{T}_{\mu \nu} \equiv \tilde{T}_{\mu \nu} +  \theta_{\mu \nu},
\end{align}
where $\tilde{T}_{\mu\nu}$ is known as the seed source and $\theta_{\mu\nu}$ an extra unknown source which introduces new physical content beyond the given by the seed matter. 

%
{In canonical coordinates, the space--time metric of a spherically symmetric and static spacetime, is given by
\begin{equation}
ds^{2}=
-e^{\nu}\,dt^{2}+e^{\lambda }\,dr^{2}
+r^{2}d\Omega^{2},
\label{metric}
\end{equation}
with $d\Omega^{2}\equiv d\theta^{2}+\sin ^{2}\theta \,d\phi ^{2}$ the usual line element of a unitary two sphere and $\nu$ and $\lambda$ are functions of the radial coordinate only. Taking into account that this general line element  in (\ref{metric}) solves the Einstein field equations 
\begin{align}
\label{EinEqFull}
G_{\mu\nu}\equiv R_{\mu \nu} - \frac{1}{2}g_{\mu \nu} R = 8\pi {T}_{\mu \nu},
\end{align}
one arrives at
\begin{align}
\label{ec1}
-8\pi \rho &= -\frac{1}{r^2}+e^{-\lambda}\left(\frac{1}{r^2}-\frac{\lambda'}{r}\right)\,,
\\\label{ec2}
8\pi {p}_{r}&=- \frac{1}{r^2}+e^{-\lambda}\left(\frac{1}{r^2}+\frac{\nu'}{r}\right)\,,
\\\label{ec3}
8\pi {p}_{\perp} &=\frac{1}{4}e^{-\lambda}\bigg(2 \nu ''+ \nu'^{2}-\lambda ' \nu' 
+ 2 \frac{\nu'-\lambda'}{r}\bigg),
\end{align}
where primes denote derivation with respect to the radial coordinate, $r$, and the quantities $\{{\rho}, {p}_{r}, {p}_{\perp}\}$ refer to the effective density and to the effective pressures along the radial and tangential directions, respectively. 

{
The seed energy-momentum tensor can be cast as
$\tilde{T}_{\mu}^{\nu}=\text{diag}\{ -\tilde{\rho}, \tilde{p}_r, \tilde{p}_{\perp},\tilde{p}_{\perp} \}$, while the unknown generic fluid, the so-called decoupling fluid as $\theta_{\mu}^{\nu} = \text{diag}\{ \theta_{0}^{0}, \theta_{1}^{1}, \theta_{2}^{2}, \theta_{3}^{3} \}$. In this way the effective thermodynamic quantities are
\begin{eqnarray}  \rho&=&\tilde{\rho}-\theta^{0}_{0}, \\
    p_{r}&=&\tilde{p}_{r}+ \theta^{1}_{1},\\  p_{\perp}&=&\tilde{p}_{\perp}+\theta^{2}_{2}.
\end{eqnarray}
}

Next, after implementing the following splitting in the metric potentials
\begin{eqnarray}\label{expectg}
\nu&=&\xi+ \alpha g \label{expectg1}\\
e^{-\lambda}&=&\mu+\alpha f, 
\end{eqnarray}
the system (\ref{ec1}) decouples in two set of differential equations: one
for the metrics $\xi$ and $\mu$ sourced by $\tilde{T}_{\mu\nu}$ and other for $f$ and $g$ sourced by $\theta_{\mu \nu}$. 
\\

In this work, we will focus on the MGD, which corresponds to setting $g=0$ such that $g_{tt}=e^{\nu}$ is always the same, and all the effects coming from the decoupling--sector lie on the radial metric potential \cite{Ovalle:2017fgl,Ovalle:2017wqi}.} {Now, plugging the transformations (\ref{expectg}) into (\ref{ec1})--(\ref{ec3}) under the mentioned considerations, one gets
\begin{align}
\label{ec1pf}
8\pi\tilde{\rho}
&=\frac{1}{r^2}-\frac{\mu}{r^2}-\frac{\mu'}{r}\ ,
\\
\label{ec2pf}
8\pi
\tilde{p}_{r}
&=
-\frac 1{r^2}+\mu\left( \frac 1{r^2}+\frac{\xi'}r\right)\ ,
\\
\label{ec3pf}
8\pi
\strut\displaystyle
\tilde{p}_{\perp}
&=
\frac{\mu}{4}\left(2\xi''+\xi'^2+\frac{2\xi'}{r}\right)
+\frac{\mu'}{4}\left(\xi'+\frac{2}{r}\right)
\ ,
\end{align}
along with the conservation equation 
\begin{equation}
\label{conpf}
\frac{d{\tilde{p}}_{r}}{dr}+\frac{\xi'}{2}\left({\tilde{\rho}}+{\tilde{p}}_{r}\right)-\frac{2}{r}\left({\tilde{p}}_{\perp}-{\tilde{p}}_{r}\right)=0,
\end{equation}
and
\begin{align}
\label{ec1d}
8\pi\,\theta_{0}^{0}
&=
\strut\displaystyle \alpha\frac{f}{r^2}
+\alpha\frac{f'}{r}\ ,
\\
\label{ec2d}
8\pi
\strut\displaystyle
\,\theta_{1}^{1}
&= \alpha f\left(\frac{1}{r^2}+\frac{\xi'}{r}\right)\ ,
\\
\label{ec3d}
8\pi
\strut\displaystyle\,\theta_{2}^{2}
&=\frac{\alpha f}{4}\left(2\xi''+\xi'^2+2\frac{\xi'}{r}\right)
+\frac{\alpha f'}{4}\left(\xi'+\frac{2}{r}\right)
\ ,
\end{align}
with 
\begin{equation}
\label{con1d}
\left(\theta_{1}^{1}\right)'-\strut\displaystyle\frac{\xi'}{2}\left(\theta_{0}^{0}-\theta_{1}^{1}\right)-\frac{2}{r}\left(\theta_{2}^{2}-\theta_{1}^{1}\right) = 0.
\end{equation}
}

At this point, some comments are in order. First, note that Eqs. (\ref{ec1pf})--(\ref{ec3pf}) solve the system when $\alpha=0$. In this regard, this set corresponds to identities after we provide a well-known seed, $\tilde{T}_{\mu\nu}$, which could be a perfect fluid, a charged fluid, etc. Second, as the $g_{tt}$ component of the metric is the same for each sector, the set 
(\ref{ec1d})-(\ref{ec3d}) corresponds to three differential equations with four unknowns, namely $\{f,\theta^{0}_{0},\theta^{1}_{1},\theta^{2}_{2}\}$ so extra information is required to close the system. This condition could be, for example, a well-motivated relation among the $\theta$--sector components (a type of equation of state), or some constraint on the spacetime scalar curvature, to name a few. Finally, it can be shown that both matter sectors are separately conserved, namely
\begin{eqnarray}
\nabla_{\mu}\tilde{T}^{\mu\nu}&=&0\\
\nabla_{\mu}{\theta}^{\mu\nu}&=&0,
\end{eqnarray}
which entails that there is no exchange of energy between them, which resembles how dark matter interacts with baryonic matter.

\section{Non-Schwarzschild black holes}\label{section3}

As stated at the end of the previous section, to close the $\theta$-sector, Eqs. (\ref{ec1d})-(\ref{ec3d}), one needs to specify both the seed spacetime and an additional constraint imposed on the components of the $\theta_{\mu\nu}$ tensor. In this context, here we consider the seed solution to be the well-known Schwarzschild BH solution whose line element reads
 \begin{equation}\label{RN}
    ds^{2}=-hdt^{2}+h^{-1}dr^{2}+r^{2}d\Omega^{2}, 
 \end{equation}
 where 
 \begin{equation}\label{SeedMetric}
     h=e^{\xi}=\mu=1-\frac{2M}{r},
 \end{equation}
with $M$ being the mass parameter. The Schwarzschild solution is Ricci flat as by the energy-momentum tensor $\tilde{T}_{\mu\nu}=0$. Therefore, in applying the MGD protocol, we are going to obtain a non-Schwarzschild metric, that is, a spacetime having $g_{tt}g_{rr}\neq -1$. Therefore, whether this remains or not being a BH spacetime strongly depends on the additional piece usually introduced by MGD on the radial metric potential. This is so because the original causal structure could be destroyed by the appearance of new critical points, which can be coordinate or physical singularities. Therefore, it is very relevant to endow the $\theta$-sector with physical meaning to avoid any undesirable situation. 

\subsection{The scalar-vector action}

The solution of the system (\ref{ec1d})-(\ref{ec3d}) will be carried out by considering that the $\theta$-sector is described by a scalar-vector interaction. The action representing this interaction is given by
\begin{equation}\label{sva}
S=\int \mathrm{d}^4 x \sqrt{-g} \big[ X^m - F^s V(\phi) \big],
\end{equation}
where, for the scalar field, we consider the kinetic term to be of the form
\begin{equation}\label{kinetic}
X = -\frac{1}{2} \nabla_\mu \phi \nabla^\mu \phi,
\end{equation}
and $V(\phi)$ is a generic function depending on the scalar field. For the vector field one has
\begin{equation}
F=-\frac{1}{4} F_{\mu \nu} F^{\mu \nu}, \quad F_{\mu \nu}=\nabla_\mu A_\nu-\nabla_\nu A_\mu.
\end{equation}
{It is worth noting that the matter sector considered here contains two distinct contributions: a scalar kinetic term $X^m$ and a nonlinear vector contribution $F^s V(\phi)$. In principle, the model contains both scalar and vector sectors; however, due to their coupling, they cannot be treated independently. The scalar-only case would correspond to a k-essence type source governed by the kinetic structure $X^m$, while the vector-only case would reduce to a nonlinear electrodynamic theory. In the present work, we focus on the combined scalar--vector interaction, since such couplings naturally arise in effective field theory extensions of gravity and provide a convenient closure of the decoupling sector within the MGD framework.}

Here, the exponents $m$ and $s$ are positive real numbers. The action (\ref{sva}) can be seen as a particular case of the action provided in \cite{Dehghani:2019cuf}, corresponding to a non-minimal coupling between scalar field and non-linear electrodynamics.

Next, the full set of equations (\ref{ec1d})-(\ref{ec3d}) reads as follows

\begin{widetext}
\begin{equation}
\begin{aligned}\label{eq25}
 X^m+(2 s - 1) F^s V(\phi)= & \strut\displaystyle \frac{\alpha}{8\pi}\left[\frac{f}{r^2}
+\frac{f'}{r}\right]\ , \\
(1-2m) X^m +(2 s - 1) F^s V(\phi)= & \frac{\alpha}{8\pi}f\left(\frac{1}{r^2}+\frac{\xi'}{r}\right)\ ,
\\ 
 X^m - F^s V(\phi)= & \frac{\alpha}{8\pi}\left[\frac{f}{4}\left(2\xi''+\xi'^2+2\frac{\xi'}{r}\right)
+\frac{f'}{4}\left(\xi'+\frac{2}{r}\right)\right].
\end{aligned}
\end{equation}
\end{widetext}

along with the conservation equation
\begin{equation}\label{eq26}
\begin{split}
  m X^{m-1} \left[ (2 m - 1 )\phi'' + \left(\frac{2}{r}+\frac{\nu ' + (1-2m)\lambda '}{2}\right) \phi'\right] &\\ - e^{\lambda} F^s\dfrac{d V}{d \phi} + \dfrac{s e^{\lambda} F^s V}{\phi'} \dfrac{d}{d r} \left[ \ln \left( r^4 V^2 F^{2s -1} \right) \right] = 0.
  \end{split}
\end{equation}
Besides, the equations of motion for the scalar field and for the vector field are
\begin{equation}\label{eq27}
    (2 m - 1 )\phi'' + \left(\frac{2}{r}+\frac{\nu ' + (1-2m)\lambda '}{2}\right) \phi' = \dfrac{1}{m} \dfrac{e^{\lambda} F^s}{X^{m-1}}\dfrac{d V}{d \phi},
\end{equation}

\begin{equation}\label{eq28}
    r^4 V^2 F^{2s -1} = \text{const},
\end{equation}
respectively. It is worth mentioning that Eqs. (\ref{eq27}) and (\ref{eq28}) are contained by the conservation Eq. (\ref{eq26}). Moreover, since Eq. (\ref{eq26}) is a linear combination of the field equations (\ref{eq25}), so no new information can be obtained from the it, and the set (\ref{eq25}) must be supplemented with extra information to determine $\{f,\phi,V,A_{\mu}\}$.

\subsection{The metric}\label{subsec:metric}

The closure of the $\theta$-sector is not only subject to its identification with a well-motivated matter content, but also it is necessary to consider some functional relation between its components, such as an equation of state (EoS). In this regard, we consider a general relation given by
\begin{equation}\label{eos}
\theta_0^0=a \theta_1^1+b \theta_2^2,
\end{equation}
where $a$ and $b$ are positive dimensionless constants. 
In this way, the equation (\ref{eos}) provides an interesting relation among the parameters $\{a,b,s,m\}$ and the scalar and vector fields
\begin{equation}
    (-2 a s+a+b+2 s-1) F^s V(\phi) - (-1 + a + b - 2 a m) X^m = 0.
\end{equation}
The above relation is valid if and only if the exponents satisfy
\begin{equation}
    m = \frac{a+b-1}{2 a}, \quad s = \frac{a+b-1}{2 (a-1)} .
\end{equation}
On the other hand, the geometric part leaves the following differential equation for the decoupler function $f(r)$
\begin{widetext}
\begin{equation}
   \frac{d\ln f(r)}{dr}=\frac{2\left[\left(2a+b-4\right)Mr-(a-1)r^{2}-(b-4)M^{2}\right]}{\left[2M-r\right]\left[\left(b-4\right)M-\left(b-2\right)r\right]r},
\end{equation}
\end{widetext}
leading to the following result
\begin{equation}\label{decoupler}
    f(r)=\frac{\left[r-2M\right]\left[\left(4-b\right)M+\left(b-2\right)r\right]^{c}}{r}l,
\end{equation}
where 
\begin{equation}\label{eq34}
    c\equiv \frac{2(1-a)}{(b-2)},
\end{equation}
and $l$ an integration constant with units of $\text{length}^{-c}$.

At this point, some comments are in order. First, the decoupler function (\ref{decoupler}) contains the seed metric potential. Therefore, the full inverse radial metric component reads as follows 
\begin{equation}\label{MGDmetric}
\begin{split}
    g^{-1}_{rr}(r)=\left[1-\frac{2M}{r}\right]\left[1+l\alpha\left(r\left(b-2\right)-\left(b-4\right)M\right)^{c}\right].
    \end{split}
\end{equation}
In this way, the minimally deformed Schwarzschild BH is given by the following line element

\begin{equation}\label{MGDmetric1}
\begin{split}
    ds^{2}=-\left[1-\frac{2M}{r}\right]dt^{2}+\left[1-\frac{2M}{r}\right]^{-1}\\ \times\left[1+l\alpha\left(r\left(b-2\right)-\left(b-4\right)M\right)^{c}\right]^{-1}dr^{2}+r^{2}d\Omega^{2}.
    \end{split}
\end{equation}

Now, a well established BH solution should preserve the Lorentzian signature throughout the whole spacetime. Furthermore, as we are interested in saving the asymptotic behavior of the seed manifold, it is imperative to restrict the new information coming from the decoupler sector. To avoid changes in signature, it is necessary to ensure that no new horizons are formed after introducing the $\theta$-sector. This is so because, once the new horizon $r^{*}_{H}$, provided by the condition $g^{-1}_{rr}(r)=0$ is crossed, only the radial metric component changes sign producing the following signature $\{-,-,+,+\}$, assuming $r^{*}_{H}>2M$. Otherwise, if $0<r^{*}_{H}<2M$ then, one obtains an Euclidean signature, that is, 
$\{+,+,+,+\}$. Therefore, in both cases, the spacetime signature is ill-defined. Then,
\begin{equation}
    \left[1-\frac{2M}{r}\right]\left[1+l\alpha\left(r\left(b-2\right)-\left(b-4\right)M\right)^{c}\right]=0,
\end{equation}
provides 
\begin{equation}
    r_{1}=2M, \quad r_{2}=\frac{(b-4)M+(-l\alpha)^{-1/c}}{b-2}
\end{equation}

The condition of avoiding the introduction of an additional causal structure beyond the original is to consider $l\alpha>0$. In such a case, the additional $r_{2}$ roots are imaginary roots. On the other hand, for negative $l\alpha$, the only chance of having just one horizon (event horizon, Killing horizon), is for those values of $l\alpha$ satisfying 
\begin{equation}\label{eq38}
    -(4M-bM)^{-c}<l\alpha<0,
\end{equation}
and also for those values of parameter $b$ satisfying
\begin{equation}\label{eq39}
    2<b<4.
\end{equation}

{Therefore, we have two branches, the branch one (B1): $l\alpha>0$ and the branch two (B2): $l\alpha<0$ leading to BH solutions.}

\begin{figure}
    \centering     \includegraphics[scale=0.6]{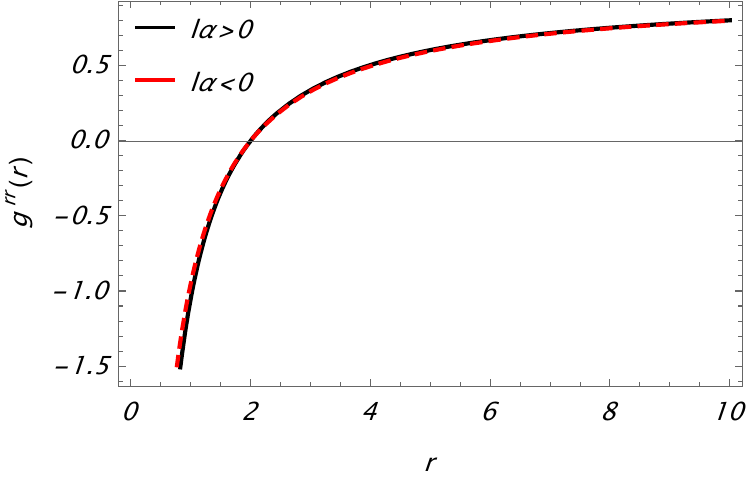}
        \caption{The inverse radial metric potential versus the radial coordinate for $M=1$, $b=3$ and $c=-2$. The B1 corresponds to the solid black line while the red dashed line corresponds to B2. in both cases we used $\alpha=0.2$ and $l=1$ and $l=-1$, respectively. }  
    \label{metricpotentialplot}
\end{figure}

As it is desirable to keep the asymptotic behavior of the seed solution, the parameter $c$ defined by Eq. (\ref{eq34}) should be negative in nature. This is compatible with conditions (\ref{eq38}) and (\ref{eq39}) {for B2}. In Fig. \ref{metricpotentialplot} it is shown the trend of the inverse radial metric potential against the radial coordinate, for both branches, where it is clear that the solution has only one horizon. {These restrictions guarantee that the deformation does not introduce additional real roots of the radial metric component outside the Schwarzschild horizon, thus preserving the original causal structure of the seed geometry.}

\begin{figure}
    \centering     \includegraphics[scale=0.46]{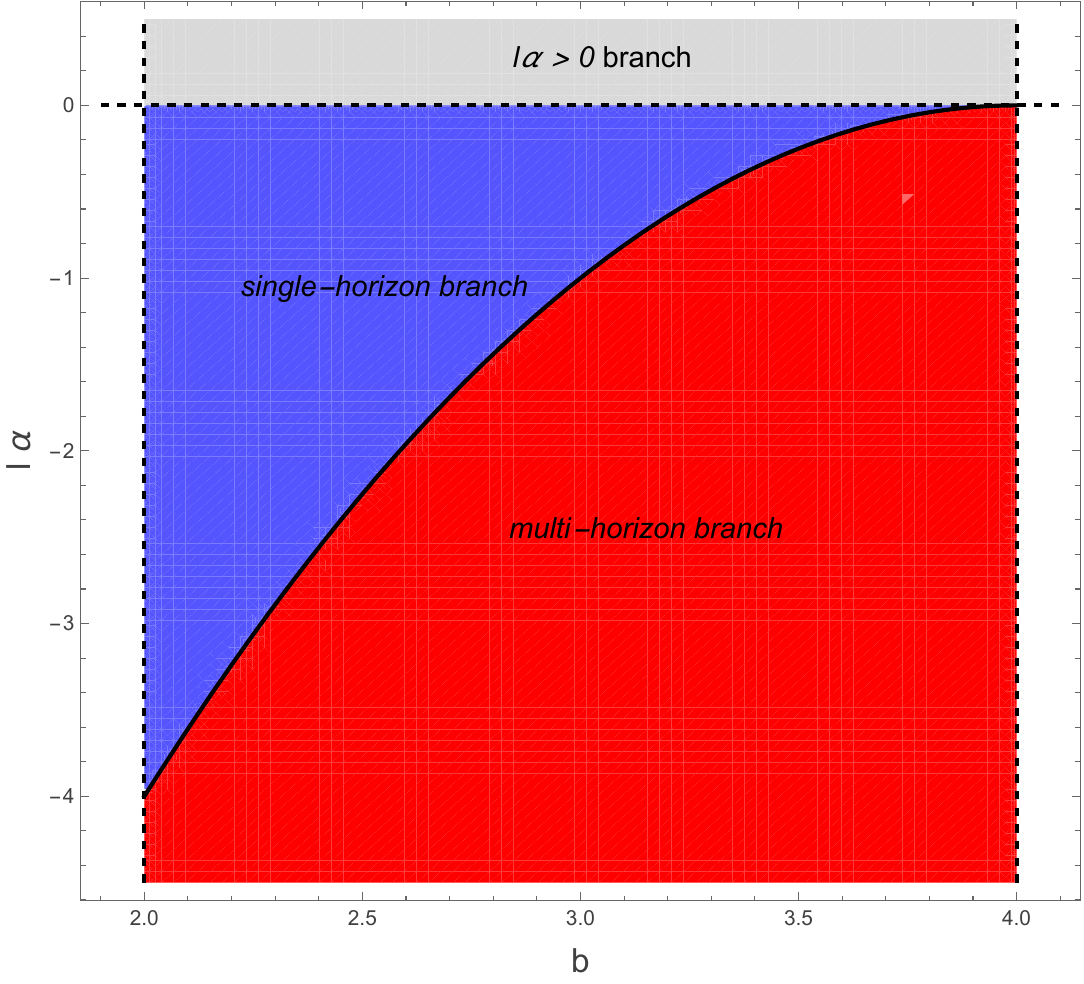}
        \caption{
Parameter space structure of the minimally deformed solution in the $(b,l\alpha)$ plane for B2 in the representative case $M=1$ and $c=-2$. The solid curve $l\alpha_{\rm crit}(b)=-(4-b)^2$ separates the single–horizon branch (blue) from the multi–horizon branch (red). The dashed horizontal line $l\alpha=0$ corresponds to the Schwarzschild limit where the deformation vanishes, while the gray region above it corresponds to B1. The vertical dashed lines indicate the interval $2<b<4$ required for the regular behavior of the deformation.
}  
    \label{spaceparemeter}
\end{figure}

{The above conditions define a restricted region in the parameter space {for the branch $l\alpha<0$} where the resulting geometry represents a BH spacetime. In particular, the combined requirements of asymptotic flatness, Lorentzian signature, and the absence of additional horizons constrain the parameters $(b,l\alpha)$ to lie within a finite domain.}
{
To illustrate these constraints, in Fig. \ref{spaceparemeter} we display the allowed region in the parameter space $(b,l\alpha)$ for $c=-2$ and $M=1$.
The horizontal axis represents  the parameter $b$, while the vertical axis corresponds to $l\alpha$. The solid black line BH denotes the critical boundary 
\begin{equation}
l\alpha_{\rm crit}(b)=-(4-b)^2M^{2},
\end{equation}
obtained from the condition that the second factor of the radial metric component becomes zero. This curve separates two qualitatively distinct branches of BH solutions. 
The blue shaded region above the curve corresponds to the \emph{single–horizon branch}, where the only root of $g_{rr}^{-1}$ is located at $r=2M$, corresponding to the Schwarzschild horizon. In this region, the MGD deformation preserves the causal structure of the Schwarzschild seed geometry, and no additional horizons are generated. 
The red shaded region below the critical curve corresponds to the \emph{multi–horizon branch}. In this case, the deformation term introduces an additional root of the radial metric function, producing a spacetime structure with more than one horizon. 
The dashed horizontal line represents $l\alpha=0$, which corresponds to the Schwarzschild limit where the deformation vanishes. The gray region above this line corresponds to B1. Finally, the vertical dashed lines indicate the interval \eqref{eq39}, which follows from the conditions required to maintain the regular behavior of the deformation function and the existence of well–defined horizons.} {Moreover, for B2 if $b<2$ the resulting spacetime becomes a wormhole structure. However, we are not going to discuss this type of solutions here.   }

Next, as the field equations (\ref{eq25}) are quite involved, it is not easy in general to determine either the scalar field or the function $V$, and in principle, the electromagnetic vector field cannot be easily determined either. However, in the representative branch, it can be done (see below). Now, the expression for the derivative of the scalar field is obtained by subtracting the first and second equations in the set (\ref{eq25}). So, this provides

\begin{widetext}
\begin{equation}\label{scalarfeldfull}
    \phi'(r)^2 = -\frac{r ((b-2) r-(b-4) M)^{\frac{2 a}{b-2}} \left(\frac{l\alpha (a-1) 2^m (2 M-r) ((b-2) r-(b-4) M)^{-\frac{2 a+b-4}{b-2}}}{m r^2}\right)^{1/m}}{(r-2 M) \left(((b-2) r-(b-4) M)^{\frac{2 a}{b-2}}+l\alpha ((b-2) r-(b-4) M)^{\frac{2}{b-2}}\right)}.
\end{equation}
\end{widetext}
The above expression blows up at the event horizon. Therefore, to avoid any pathological behavior, we shall consider $m=1$. 
Now, using the third equation of \eqref{eq25}, one obtains

\begin{widetext}
    \begin{equation}\label{eq42}
    - s F^{s-1} V(\phi) = \alpha  \frac{2^{\frac{a+b-1}{2 (a-1)}} ((b-2) r-(b-4) M)^{\frac{2}{b-2}-1} (M (-3 a+b-1)+r (a-b+1))}{r^2 A^{'2}_{0} \left(((b-2) r-(b-4) M)^{\frac{2 a}{b-2}}+l\alpha  ((b-2) r-(b-4) M)^{\frac{2}{b-2}}\right)}.
\end{equation}
\end{widetext}

{As can be observed, the previous expression depends on $A'_0$, that is, the derivative of the temporal component of the four-potential $A_\mu$. This comes from the spherical and static features of the model, where there is only one non-trivial component. In connection with the vector field equation \eqref{eq28}, in principle it can be formally integrated to obtain the temporal component of the vector potential $A_\mu$, however, one needs to specify the form of the coupling function $V$ (see below). Therefore, we can at least express the invariant $F$ as
\begin{equation}\label{F}
    F(r)=\left[\frac{Q^2}{2r^4V^2}\right]^{\frac{1}{2s-1}},
\end{equation}
where the integration constant has been interpreted as an electric charge. 
The integration constant $Q$ appearing in the above expression does not correspond to the standard Maxwell electric charge. Instead, in nonlinear electrodynamics it is associated with the conserved flux of the electromagnetic displacement tensor that follows from the field equations of the nonlinear electromagnetic sector.
Due to the nonlinear electromagnetic coupling, this quantity does not manifest as an asymptotic gravitational charge constant associated with the vector field. To better understand this point, let us expand the radial metric potential, that is,
\begin{equation}
\begin{split}
    g_{rr}(r)=\left[1+\frac{2M}{r}+\mathcal{O}\left(\frac{1}{r^2}\right)\right]\bigg[1-\frac{l\alpha}{\left(b-2\right)^2r^2}\\+\mathcal{O}\left(\frac{1}{r^4}\bigg)\right],
    \end{split}
\end{equation}
where we have considered $c=-2$. Therefore, one has the same asymptotic behavior as the Schwarzschild BH\footnote{Of course, a different value for the $c$ parameter could change in principle the asymptotic behavior of the solution, and here, we took the value which guarantees flatness.}  
\begin{equation}
    g_{rr}(r)=1+\frac{2M}{r}+\mathcal{O}\left(\frac{1}{r^2}\right),
\end{equation}
where it is clear that there is no evidence of the scalar and vector fields for far enough distances, both represented by the parameters $l$ and $\alpha$ due to the non-minimal coupling between them. In other words, the solution is screened by the Schwarzschild BH. Furthermore, while the action contains a nonlinear electromagnetic sector, the anisotropic closure employed to obtain the exact solution prevents a simultaneous Maxwell–canonical scalar limit. Therefore, the present geometry should be interpreted as a distinct branch rather than a continuous deformation of Einstein–scalar–Maxwell BH.}


\subsection{Explicit matter-sector reconstruction}

Although the general system determining the scalar–vector sector does not
admit a simple closed-form solution for arbitrary parameters, the situation
simplifies considerably in the representative parameter space used throughout the numerical analysis of this work. In particular, for the choice
$\{M,b,c\}=\{1,3,-2\}$,
which lies within the one-horizon region of the parameter space, the matter sector can be explicitly reconstructed.
For these values the closure relation fixes $\{a,m,s\}=\{2,1,2\}$,
so that the scalar–vector Lagrangian reduces to the particularly simple form
\begin{equation}
\mathcal{L}_{\rm m}=X - F^{2} V(\phi).
\end{equation}

\paragraph{Scalar field.}

For the parameter space $\{M,m,c,b\}=\{1,1,-2,3\}$ the expression \eqref{scalarfeldfull} reduces to 

\begin{equation}
\phi'(r)^2=
\frac{l\alpha}
{4\pi r(1+r)\left[(1+r)^2+l\alpha\right]}.
\label{phi_prime_rep}
\end{equation}

Since all factors in the denominator are positive for
$r\geq r_H=2$ and B1, the scalar field remains real throughout
the entire exterior region. 

The equation (\ref{phi_prime_rep}) can be exactly integrated. Introducing the
variable

\begin{equation}
u=\sqrt{\frac{r}{1+r}},
\end{equation}

the scalar profile becomes

\begin{equation}
\phi(r)=
\phi_0
\pm
\sqrt{\frac{l\alpha}{\pi}}
\int
\frac{du}
{\sqrt{1+l\alpha\left(1-u^2\right)^2}},
\end{equation}

whose closed form is given by an incomplete elliptic integral of the first
kind,

\begin{equation}
\phi(r)=
\phi_0
\pm
\sqrt{\frac{l\alpha}{\pi}}
\,
\frac{1}
{\sqrt{1+l\alpha}\sqrt{p}}
\mathcal{F}\!\left[
\arcsin
\left(
\sqrt{p}
\sqrt{\frac{r}{1+r}}
\right),
\frac{q}{p}
\right],
\label{phi_exact}
\end{equation}

where

\begin{equation}
p=
\frac{l\alpha+i\sqrt{l\alpha}}
{1+l\alpha},
\qquad
q=
\frac{l\alpha-i\sqrt{l\alpha}}
{1+l\alpha},
\end{equation}

and $\mathcal{F}(\varphi,m)$ denotes the incomplete elliptic integral of the first
kind. Although Eq.~(\ref{phi_exact}) is written in terms of complex
parameters, the resulting scalar field is real for B1.

Evaluating Eq.~(\ref{phi_prime_rep}) at the event horizon
$r_H=2$ gives

\begin{equation}
\phi^{\prime 2}(r_H)=
\frac{l\alpha}
{24\pi\left(9+l\alpha\right)},
\end{equation}

which is finite over the entire admissible parameter range.
Therefore, the solution possesses a regular scalar field at
the horizon.

For large radial distances,

\begin{equation}
\phi(r)=
\phi_\infty
-
\frac{\sqrt{l\alpha}}
{\sqrt{\pi r}}
+
\mathcal{O}\!\left(r^{-3/2}\right),
\end{equation}

showing that the scalar field approaches a constant asymptotically while
its radial derivative vanishes at infinity.

It is worth emphasizing that the complementary branch
B2 exhibits a fundamentally different behavior. In this case,
Eq.~(\ref{phi_prime_rep}) becomes negative throughout the exterior region,
implying that the scalar field is no longer real unless it is interpreted
as a phantom field through the re-definition
$\phi\rightarrow i\phi$. For this reason, the scalar reconstruction
presented in this work naturally favors B1.

\paragraph{Vector sector and nonlinear coupling.}

Once the scalar sector has been reconstructed, the remaining field equations
can be used to determine the nonlinear electromagnetic sector. In particular,
substituting the explicit scalar profile into the third equation of the set
(\ref{eq25}) leads to the following relation between the electromagnetic
invariant and the scalar coupling function,

\begin{equation}
F(r)^2V(r)=
\frac{
l\alpha\left[(r+1)^2(3r-1)+l\alpha(r-1)\right]
}
{8\pi r^2(r+1)^3\left[(r+1)^2+l\alpha\right]}.
\label{FV_relation}
\end{equation}

Solving Eqs.~(\ref{F}) and
(\ref{FV_relation}) simultaneously, one obtains the
electromagnetic invariant

\begin{equation}
F(r)=
\frac{
l^{2}\alpha^{2}
\left[(r+1)^2(3r-1)+l\alpha(r-1)\right]^2
}
{
32\pi^{2}Q^{2}
(r+1)^6
\left[(r+1)^2+l\alpha\right]^2
},
\label{InvariantF}
\end{equation}

while the scalar--vector coupling function becomes

\begin{equation}
V(r)=
\frac{
128\pi^{3}Q^{4}
(r+1)^9
\left[(r+1)^2+l\alpha\right]^3
}
{
l^{3}\alpha^{3}
r^{2}
\left[(r+1)^2(3r-1)+l\alpha(r-1)\right]^3
}.
\label{PotentialV}
\end{equation}

Therefore, the interaction potential is completely reconstructed once the
geometry has been specified, and no additional ansatz for $V(\phi)$ is
required.

At large radial distances one finds

\begin{equation}
F(r)=
\frac{9l^{2}\alpha^{2}}
{32\pi^{2}Q^{2}}
\frac{1}{r^{6}}
+\mathcal{O}(r^{-7}),
\end{equation}

whereas

\begin{equation}
V(r)=
\frac{128\pi^{3}Q^{4}}
{27l^{3}\alpha^{3}}
\frac{1}{r^{2}}
+\mathcal{O}(r^{-3}).
\end{equation}

Hence, the nonlinear electromagnetic invariant decays much faster than the
coupling function, ensuring that the vector sector becomes asymptotically
suppressed. Consequently, both the scalar and nonlinear electromagnetic
fields vanish sufficiently rapidly at large distances, consistently with the
Schwarzschild asymptotics of the spacetime.

\paragraph{Electric potential.}

For a purely electric configuration
\[
A_\mu=(A_0(r),0,0,0),
\]
the electromagnetic invariant satisfies

\begin{equation}
F=
\frac12
\left(
1+\frac{l\alpha}{(r+1)^2}
\right)
A_0^{\prime 2}(r).
\end{equation}

Substituting the reconstructed invariant
(\ref{InvariantF}) yields

\begin{equation}
A_0'(r)=
\frac{
l\alpha
\left[(r+1)^2(3r-1)+l\alpha(r-1)\right]
}
{
4\pi Q
(r+1)^2
\left[(r+1)^2+l\alpha\right]^{3/2}
},
\label{A0prime}
\end{equation}

which can be integrated numerically to obtain the electric potential
$A_0(r)$.

At large distances one finds

\begin{equation}
A_0'(r)=
\frac{3l\alpha}
{4\pi Q}
\frac1{r^2}
+\mathcal O(r^{-3}),
\end{equation}

and therefore

\begin{equation}
A_0(r)=
A_{0,\infty}
-
\frac{3l\alpha}
{4\pi Q}
\frac1r
+\mathcal O(r^{-2}),
\end{equation}

showing that the electric potential approaches the standard Coulomb
behavior asymptotically.

The complete reconstruction presented above demonstrates that the
anisotropic sector generated through GD can be
consistently interpreted as an effective scalar field interacting with a
nonlinear electromagnetic field rather than representing a purely effective anisotropic fluid \cite{Cotton:2021tfl}. The scalar field remains regular
throughout the exterior region and approaches a constant at spatial
infinity, whereas the nonlinear electromagnetic invariant and the
interaction potential decay sufficiently fast so that the Schwarzschild
asymptotics of the geometry are preserved.

\subsection{Effective anisotropic source}
{Although the matter content has already been identified at the level of the action and the field profiles, it is illuminating to display the explicit components of the effective energy–momentum tensor that sources the geometry \eqref{MGDmetric1}. Since the Schwarzschild seed is Ricci-flat ($\tilde{T}_{\mu\nu}=0$), the full energy–momentum tensor reduces to $T^{\mu}{}_{\nu}= \theta^{\mu}{}_{\nu}$, and following the convention adopted below Eq.~\eqref{StressTensorEffective} we identify the effective energy density and the radial and tangential pressures as
\begin{equation}
\rho \equiv -\,\theta^{0}{}_{0}, \qquad
p_{r} \equiv\theta^{1}{}_{1}, \qquad
p_{\perp} \equiv \,\theta^{2}{}_{2}.
\end{equation}
Inserting the decoupler function \eqref{decoupler} into Eqs.~\eqref{ec1d}-\eqref{ec3d} and recalling the exponent $c$ defined in Eq.~\eqref{eq34}, a direct computation yields
\begin{align} \nonumber
\rho &= 
-\frac{l\alpha\,\big[(b-2)r-(b-4)M\big]^{\,c-1}
}{8\pi\,r^{2}}\nonumber\\
&\times \big[\,2a(r-2M)-b(r-M)\,\big], 
\\ \nonumber 
p_{r} &= 
\frac{l\alpha\,\big[(b-2)r-(b-4)M\big]^{\,c}}{8\pi\,r^{2}}, 
\\ \nonumber
p_{\perp} &= 
\frac{(a-1)\,l\alpha\,(r-M)\,\big[(b-2)r-(b-4)M\big]^{\,c-1}}{8\pi\,r^{2}}.
\end{align}
In the representative parameter space $\{a,b,c,M\}=\{2,3,-2,1\}$, used throughout the numerical analysis, these expressions reduce to the compact forms
\begin{eqnarray}\nonumber
    \rho &=& \frac{l\alpha\,(r-5)}{8\pi\,r^{2}\,(r+1)^{3}}, \\
p_{r} &=& \frac{l\alpha}{8\pi\,r^{2}\,(r+1)^{2}},\\ \nonumber
    p_{\perp} &=& \frac{l\alpha\,(1-r)}{8\pi\,r^{2}\,(r+1)^{3}}. \label{eq:components_branch}
\end{eqnarray}
The effective source is intrinsically anisotropic since
$p_r\neq p_\perp$, while all stress--energy components vanish
continuously in the Schwarzschild limit $l\alpha\rightarrow0$.
Furthermore, all components decay as $\mathcal{O}(r^{-4})$ at large
distances, consistently with the asymptotic Schwarzschild behavior of
the geometry.}

{The sign of the effective stress-energy tensor is entirely determined
by the product $l\alpha$. For the B1 case, the radial pressure remains positive throughout the
exterior region ($r\geq2$), whereas the tangential pressure is negative.
The effective energy density changes sign at $r=5$, being negative in
the interval $2\leq r<5$ and positive for $r>5$. Consequently, the weak
energy condition (WEC) is violated in the vicinity of the event horizon and in the tangential direction, while
the matter sector becomes progressively less exotic as the radial
coordinate increases. Since all components decay rapidly at infinity,
these violations remain localized in the strong-field region and do not
affect the asymptotic structure of the spacetime.}

{For completeness, the complementary branch B2 exhibits the
opposite behavior: the energy density becomes positive close to the
event horizon but negative at sufficiently large distances, whereas the
radial pressure remains negative throughout the exterior region. This
branch satisfies the odd-parity stability criterion (see below) but possesses a less
satisfactory effective matter sector \cite{Curiel:2014zba}. it is worth mentioning that, in some minimally coupled tensor-scalar scenarios, non-Schwarzschild BHs also violate the WEC \cite{Bakopoulos:2023hkh}.     }

\subsection{Stability via odd and even parity criteria}

Here we employ the master equations derived in \cite{Gannouji:2021oqz, Baez:2022rdz} for odd and even parity perturbation modes, obtained for the following theory
\begin{equation}\label{action}
S=\int \mathrm{d}^4 x \sqrt{-g}\left[f_1(\phi) \frac{R}{16\pi}+f_2(\phi, X, F)\right],
\end{equation}
where the dependency on $F$ in the function $f_{2}$ is not necessarily linear in the Maxwell field, that is, it accepts the so-called non-linear electrodynamic forms. The action (\ref{action}) corresponds to a general expression for the well-known Einstein-Maxwell-dilaton theories \cite{Gibbons:1987ps,Garfinkle:1990qj}. {The stability of BH solutions in modified gravity
theories and in the presence of additional fields has been
extensively studied in different contexts, including
$f(R)$ gravity with dynamical Chern–Simons corrections
\cite{Moon:2011fw}.}

Comparing our theory

\begin{equation}\label{eq43}
S=\int \mathrm{d}^4 x \sqrt{-g}\left[ \frac{R}{16\pi}+X^{m}-F^{s}V(\phi)\right]
\end{equation}
with (\ref{action}), one obtains 
\begin{equation}
  f_{1}(\phi)=1, \quad  f_2(\phi, X, F)=X^{m}-F^{s}V(\phi).
\end{equation}
 Thus, our proposal considers a minimal coupling between the gravitational and scalar sectors, while non-minimal coupling between the scalar and vector sectors, where in principle the vector sector corresponds to non-linear form of Maxwell interaction and the potential $V(\phi)$ is not being the usual dilatonic form. 

 Next, we are going to analyze under what conditions for the parameter space $\{m,s\}$, the odd and even parity master equations for stability are fulfilled.

\subsubsection{Odd parity constraints}

\begin{figure}
    \centering     
        \includegraphics[scale=0.6]{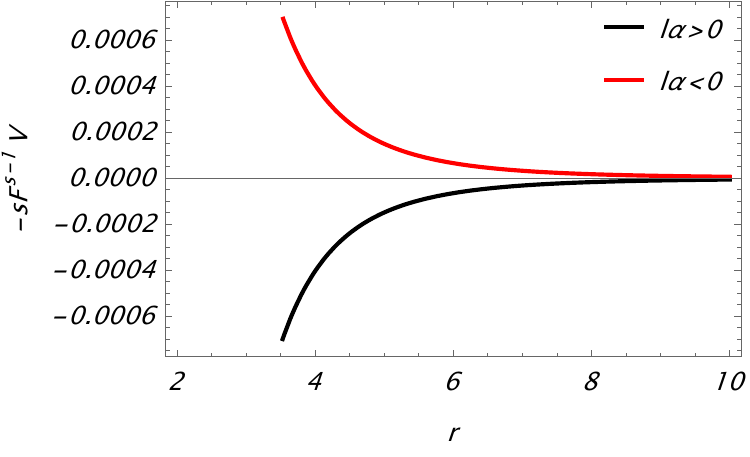}
    \caption{Odd parity condition for stability for $M=1$, $b=3$, $c=-2$ and $\alpha=0.2$, for booth cases B1 (black line) and B2 (red line).}
    \label{odd}
\end{figure}

The stability conditions imposed by the odd parity analysis entail in this case the satisfaction of the following inequalities
\begin{equation}\label{oddconstraints}
f_1(\phi)>0, \quad \frac{\partial f_{2}(\phi, X, F)}{\partial F}>0.
\end{equation}
The first one is related to the ghost-free condition for the scalar fields, while the second concerns the stability under vector perturbations. 
In our case, $f_{1}(\phi)=1$, so the first inequality is trivially satisfied. For the second condition, one gets
\begin{equation}\label{eq46}
    \frac{\partial f_{2}(\phi, X, F)}{\partial F}=-sF^{s-1}V(\phi)>0. 
\end{equation}
In this case, the odd parity stability condition is connected with the expression given by Eq. \eqref{eq42}. Then, to ensure stability via odd parity master conditions, expression \eqref{eq42} should always be positive. {This fact is shown by Fig. \ref{odd} for B1 and B2, where the condition \eqref{eq46} is satisfied only for the case B2.} 

\subsubsection{Even parity constraints}

\begin{figure}
    \centering     \includegraphics[scale=0.6]{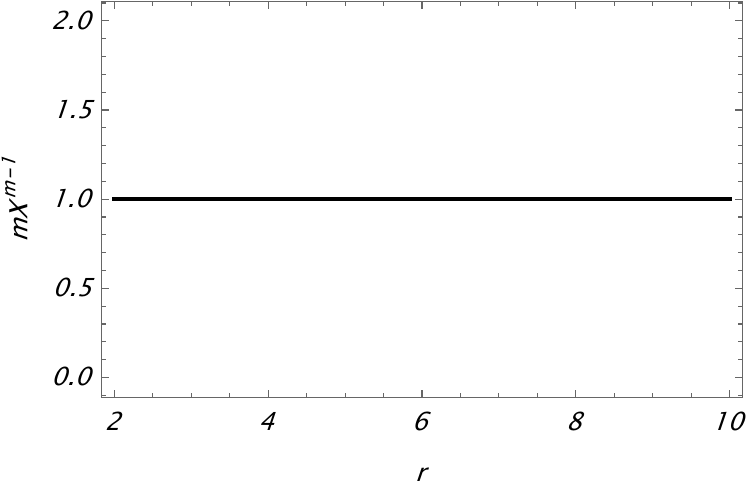}
    \caption{Even parity condition for stability for $M=1$, $b=3$, $c=-2$,  and $\alpha=0.2$, fr both cases B1 and B2. }
    \label{even}
\end{figure}

The even parity stability criterion establishes the following conditions 
\begin{equation}
f_{1}(\phi)>0, \quad 3 \left[\frac{\partial f_{1}(\phi)}{\partial \phi}\right]^{2}+2 f_1(\phi) \frac{\partial f_{2}(\phi,X,F)}{\partial X}>0.
\end{equation}
As before, the first condition is immediately satisfied, while the second one provides 
\begin{equation}\label{eq499}
    mX^{m-1}>0.
\end{equation}
It is clear that the satisfaction of the even parity stability condition is guaranteed only if $m>0$. This condition is represented in Fig. \ref{even}.

The stability analysis presented above shows that the two BH branches
possess complementary physical properties. While branch B2 satisfies both
odd- and even-parity stability criteria, branch B1 provides a more satisfactory
effective matter sector, admitting a real scalar-field reconstruction together
with a fully analytical nonlinear electromagnetic sector. Since the purpose of
the following sections is to investigate the physical properties of the matter
fields supporting the geometry, the explicit reconstruction, geodesic analysis,
and thermodynamic study will be carried out for branch B1.

\section{Geodesic motion}\label{section4}

In this section, we study the main properties of equatorial orbits around the solution described above. The geodesic motion of a test particle in a metric $g_{\mu \nu}$ is governed by the Lagrangian
\begin{eqnarray} \label{EQlagrangian}
    \mathcal{L}=\frac{1}{2} g_{\mu \nu} \Dot{x}^\mu\Dot{x}^\nu=\frac{\epsilon}{2},
\end{eqnarray}
where $\Dot{x}^\mu=d x^\mu / d\tau$ being $\tau$  an affine parameter, and $\epsilon=1$ for massive particles and $\epsilon=0$ for photons. {It is worth mentioning that, in the present analysis, we restrict ourselves to the motion of neutral test particles to isolate the purely geometrical effects associated with the MGD deformation of the radial metric component. The study of charged particle dynamics in the presence of the nonlinear electromagnetic field would require the inclusion of the Lorentz force in the equations of motion and could reveal additional dynamical features related to the scalar--vector sector. Such an analysis lies beyond the scope of the present work but represents an interesting direction for future research (see, e.g., \cite{Stuchlik:2020rls}) }

Due to the spherical symmetry, we can restrict the particle's motion to the equatorial plane ($\theta=\pi /2$, $\Dot{\theta}=0$) without loss of generality. Besides, since the metric is independent of the coordinates $t$ and $\phi$, we have two constants of motion, namely $E$ and $L$, which are the total energy and angular momentum per unit mass respectively, 
\begin{eqnarray}
    \frac{d}{d\tau}\frac{\partial \mathcal{L}}{\partial \Dot{t}}=0 &\Rightarrow& \frac{\partial \mathcal{L}}{\partial \Dot{t}}=g_{tt} \Dot{t}=E \label{EQenergy}\\
        \frac{d}{d\tau}\frac{\partial \mathcal{L}}{\partial \Dot{\phi}}=0 &\Rightarrow& \frac{\partial \mathcal{L}}{\partial \Dot{\phi}}=g_{\phi \phi} \Dot{\phi}=-L. \label{EQangularmomentum}
\end{eqnarray}

Using Eqs. (\ref{EQenergy}) and (\ref{EQangularmomentum}), we can rewrite Eq. (\ref{EQlagrangian}) as
\begin{eqnarray} \label{EQgeo1}
   - g_{tt}g_{rr} \Dot{r}^2= E^2-g_{tt} \bigg( \epsilon -\frac{L^2}{g_{\phi \phi}} \bigg).
\end{eqnarray}

\begin{figure}
    \centering     \includegraphics[scale=0.6]{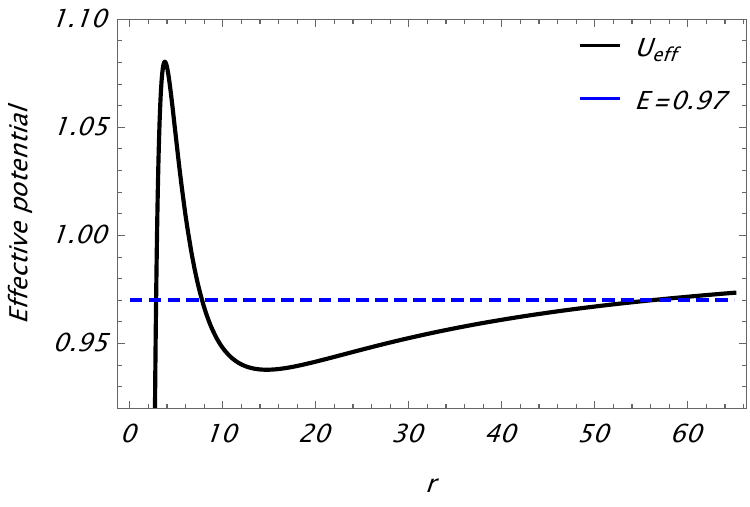}
        \caption{Radial The effective potential versus the radial coordinate for $M=1$, $b=3$, $c=-2$, $\alpha=0.2$ and $l=1$.}  
    \label{IMGpotential}
\end{figure}

Note that (\ref{EQgeo1}) is formally the same as the conservation equation in Newtonian mechanics with potential
\begin{eqnarray}
    U_{\text{eff}}=g_{tt} \bigg( \epsilon -\frac{L^2}{g_{\phi \phi}} \bigg),
\end{eqnarray}
which only depends on the metric functions $g_{tt}$ and $g_{\phi \phi}$. The effective potential $U_{\text{eff}}$ for massive particles ($\epsilon=1$) is shown in Fig. \ref{IMGpotential}. Note that a massive particle with energy $E$ and angular momentum $L$ has a bounded orbit whenever there are two intersection points between $E^2$ and $U_{\text{eff}}$ and $E^2>U_{\text{eff}}$ in the region between the intersections, as occurs in the Newtonian case.

\begin{figure*}
\centering
    \includegraphics[width=0.32\textwidth]{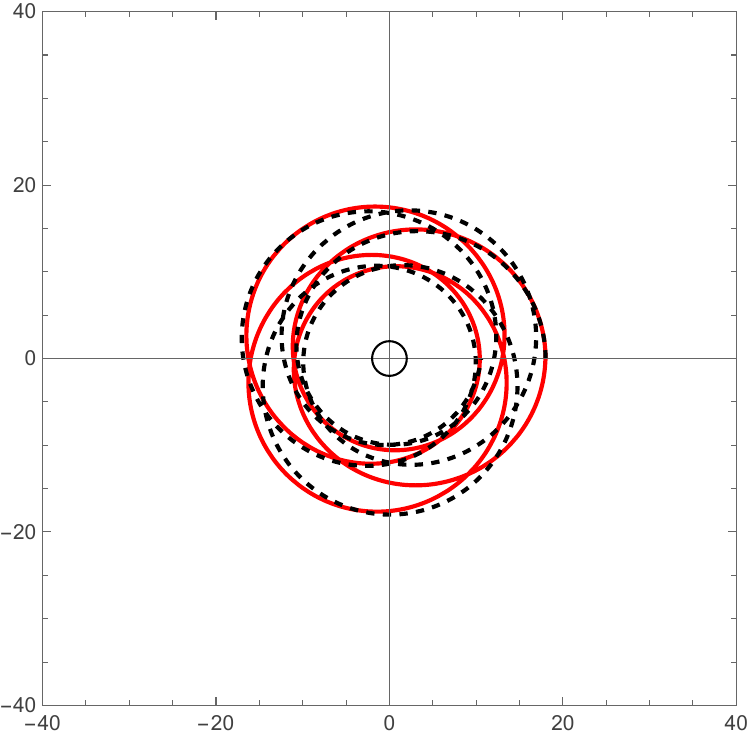} \
    \includegraphics[width=0.32\textwidth]{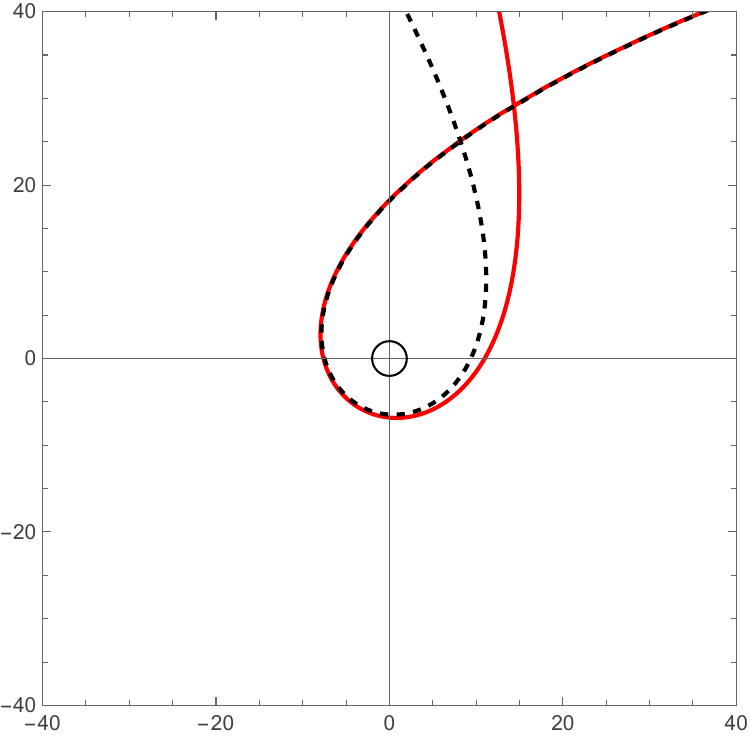}\
    \includegraphics[width=0.32\textwidth]{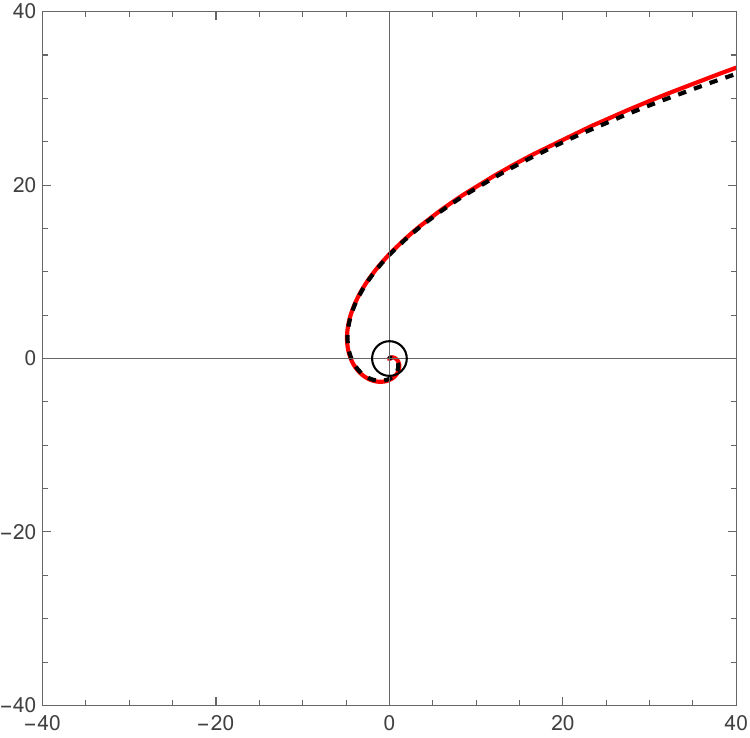}
        \caption{\textbf{Left panel}: Bounded orbits. \textbf{Middle panel}: Scattering orbit. \textbf{Right panel}: Plunge orbit. For these plots we consider $M=1$, $b=3$, $c=-2$, $\alpha=0.2$ and $l=1$. The black line corresponds to the Schwarzschild BH, and the black one to the model \eqref{MGDmetric1}.}  
    \label{orbits}
\end{figure*}


To study the geodesic motion of massive particles, we can consider
\begin{eqnarray}
    \frac{dr}{d\tau}=\frac{dr}{d\phi}\frac{d\phi}{d\tau}= -\frac{L}{g_{\phi \phi}} \frac{dr}{d\phi},
\end{eqnarray}
to rewrite Eq. (\ref{EQgeo1}) as 
\begin{eqnarray}\label{EQboundorbits}
    \frac{dr}{d\phi}&=&\sqrt{-\frac{g_{\phi \phi}^2 }{L^2 g_{rr} g_{tt}} \big( E^2 -U_{\text{eff}}\big)}\\ &=&\sqrt{-\frac{g_{\phi \phi}^2 }{L^2 g_{rr}} \bigg(\frac{E^2}{g_{tt}}+\frac{L^2}{g_{\phi \phi}}-\epsilon \bigg)},
\end{eqnarray}
from where it can be easily understood the necessity for a bound orbit to have $E^2 > U_{\text{eff}}$ such that the square root always yields real values. To solve the Eq. (\ref{EQboundorbits}), we follow the standard procedure of Chandrasekar \cite{Chandrasekhar1992} by considering the parametrization of a bounded orbit
\begin{eqnarray}
    r(\chi)=\frac{\tilde{a}}{1+\tilde{b} \cos \chi},
\end{eqnarray}
where $\tilde{a}$ and $\tilde{b}$ are constants such that $r(0)=r_{\text{min}}$ and $r(\pi)=r_{\text{max}}$ corresponding to the periastron and the apastron respectively. Then we arrive at the differential equation
\begin{eqnarray}\label{EQgeonum}
    \frac{d\phi}{d\chi}=\frac{\tilde{a}\tilde{b}L \sin \chi}{(1+\tilde{b} \cos \chi)^2} \sqrt{-\frac{g_{rr}g_{tt}}{g_{\phi \phi}^2\big( E^2 -U_{\text{eff}}\big)}},
\end{eqnarray}
that can be solved numerically. The recipe is to choose a value for $E$ and $L$ such that $E^2>U_{\text{eff}}$ has a well-defined region with a minimum and maximum radius. Here we choose $E=0.97$ and $L=4.3$. Next, we find the values of the periastron and apastron for these particular parameters, in our case $r_{\text{min}}=2.753$ and $r_{\text{max}}=19.331$ from where $\tilde{a}=4.819$ and $\tilde{b}=0.705$. It is worth noticing that the values for $r_{\text{min}}$ and $r_{\text{max}}$ will be the same for all the geodesics shown in this work. This information is enough to solve Eq. (\ref{EQgeonum}) numerically.

Figure \ref{orbits} displays three qualitatively distinct classes of timelike geodesics: 
bounded orbits (left panel), scattering orbits (middle panel), and plunge orbits (right panel). 
In all panels, the dotted black curve corresponds to the Schwarzschild solution, whereas the 
solid red curve represents the orbit in the scalar--vector geometry \eqref{MGDmetric1} for the 
same conserved quantities $E$ and $L$, using the parameters 
$M=1$, $b=3$, $c=-2$, $\alpha=0.2$, and $l=1$.

\paragraph*{Bounded orbits (left panel).}
Bounded trajectories oscillate between a periastron radius $r_{\rm min}$ and an apastron radius 
$r_{\rm max}$. Because the effective potential $U_{\rm eff}$ is identical in both 
spacetimes---a consequence of the fact that the MGD deformation leaves $g_{tt}$ and $g_{\phi\phi}$ unchanged---the radial motion coincides exactly in the two geometries. As a result, the values of $(r_{\rm min},r_{\rm max})$ 
agree identically. 
The azimuthal motion, however, depends on the radial metric component $g_{rr}$ through the relation
\begin{equation}
\frac{d\varphi}{dr} \propto \sqrt{\frac{g_{\phi\phi}^2}{L^2 g_{rr} g_{tt}}}\,
\frac{1}{\sqrt{E^2-U_{\rm eff}(r)}} ,
\end{equation}
and therefore differs between the two spacetimes.  
This produces distinct periapsis precessions: the rosette-shaped red orbit accumulates a different 
azimuthal angle per radial period when compared with the Schwarzschild one.  
The left panel thus shows that the MGD deformation breaks the exact degeneracy between the radial and 
azimuthal frequencies, generating a measurable change in the perihelion advance while preserving the 
turning points of the motion.

\paragraph*{Scattering orbits (middle panel).}
The middle panel shows unbound (scattering) trajectories.  
For the chosen energy $E$, the particle comes from spatial infinity, reaches a minimum radius 
$r_{\rm min}$---determined solely by the shape of $U_{\rm eff}$ and therefore identical in both metrics---and then 
returns to infinity.  
The total deflection angle,
\begin{equation}
\Theta = \pi - 2 \int_{r_{\rm min}}^{\infty} \frac{d\varphi}{dr}\,dr,
\end{equation}
is nevertheless sensitive to the radial metric function $g_{rr}$.  
Accordingly, the red trajectory exhibits a deflection that differs from the Schwarzschild prediction, 
despite the fact that the effective potential and the asymptotic structure are the same.  
This constitutes a clean, coordinate-invariant imprint of the MGD deformation on dynamical observables.

\paragraph*{Plunge orbits (right panel).}
The right panel now illustrates plunge trajectories for which the particle has no outer turning 
point: once it crosses the maximum of the effective potential, it inevitably falls toward the horizon.  
In this case, the red and black curves \emph{coincide almost perfectly throughout the entire trajectory}.  
This complete overlap reflects the fact that plunging trajectories probe only the radial inequality 
$E^{2}>U_{\rm eff}(r)$ and do not undergo oscillatory motion.  
Since $U_{\rm eff}$ and $g_{tt}$ are identical in the two geometries, the combination 
\begin{equation}
E^{2}-U_{\rm eff}(r)
\end{equation}
that governs the monotonic radial fall is the same.  
Although in principle the accumulated azimuthal angle before crossing the horizon depends on $g_{rr}$, 
for plunging trajectories this effect is too small to produce a visible separation of the curves.  
Consequently, the plunge orbits in Figure \ref{orbits} are practically indistinguishable from their 
Schwarzschild counterparts, in contrast with the bounded and scattering cases.


In summary, Figure \ref{orbits}  demonstrate that minimally deformed BHs 
obtained via GD share the same global orbit classification and the same radial 
turning points as Schwarzschild, but exhibit quantitative and observable differences in the 
periapsis precession and scattering angle.  
Plunge orbits constitute the only exception: due to their strictly monotonic radial behavior, they are 
effectively insensitive to the MGD corrections and therefore lie on top of the Schwarzschild results.

\section{Thermodynamic aspects}\label{section5}

In this section, we present a strengthened and conceptually sharpened analysis of the
thermodynamics associated with the minimally deformed Schwarzschild geometry introduced
earlier. 

{The thermodynamic interpretation of BHs, established through the four laws of BH mechanics \cite{Bardeen:1973gs}, the entropy-area relation \cite{Bekenstein:1973ur}, the discovery of Hawking radiation \cite{Hawking:1975vcx}, and the consolidation of BH thermodynamics \cite{Hawking:1976de}, provides the framework for the following analysis.}

As the event horizon remains fixed at $r_{H}=2M$, the parameter $M$ can be formally identified as the BH mass.  
Although the functional form of $g_{tt}$ coincides with the Schwarzschild one, the deformation encoded in $f$ fundamentally modifies the redshift properties near the horizon and is responsible for the  deviations in all thermodynamic quantities.
The parameters $\alpha$ and $l$ therefore act as control parameters, and the classical limit is easily recovered as $\alpha\to0$.

The Hawking temperature follows directly from the surface gravity associated with the
timelike Killing field. Using the general expression \cite{Angheben:2005rm}

\begin{equation}
T_H=\frac{1}{4 \pi} \sqrt{\partial_r g_{tt}\left(r_{H}\right) \partial_r g^{rr}\left(r_{H}\right)},
\label{eq:TH_final}
\end{equation}
One obtains 
\begin{equation}\label{eq58}
    T_{H}=\frac{1}{4\pi r_{H}}\sqrt{\frac{2^c+l\alpha(b r_{H})^c }{2^c}}.
\end{equation}

\begin{figure*}
\centering
    \includegraphics[width=0.32\textwidth]{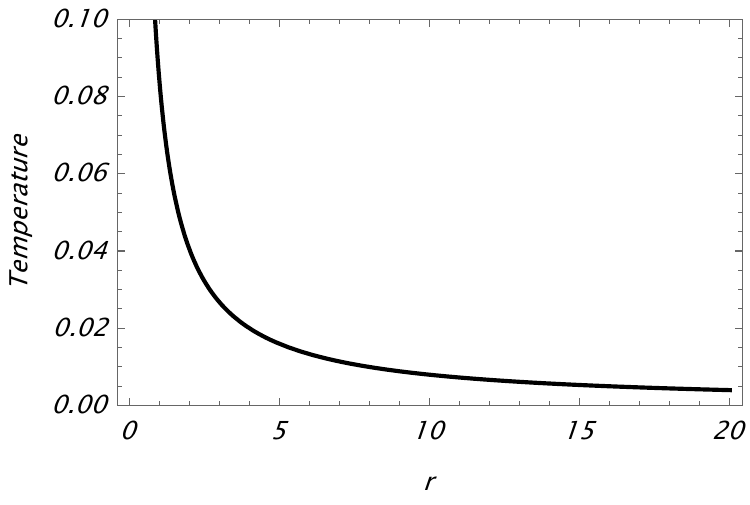} \
    \includegraphics[width=0.32\textwidth]{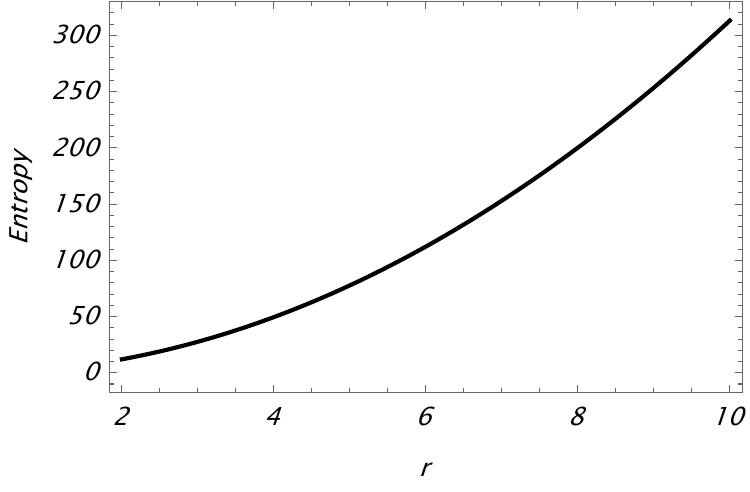}\
    \includegraphics[width=0.32\textwidth]{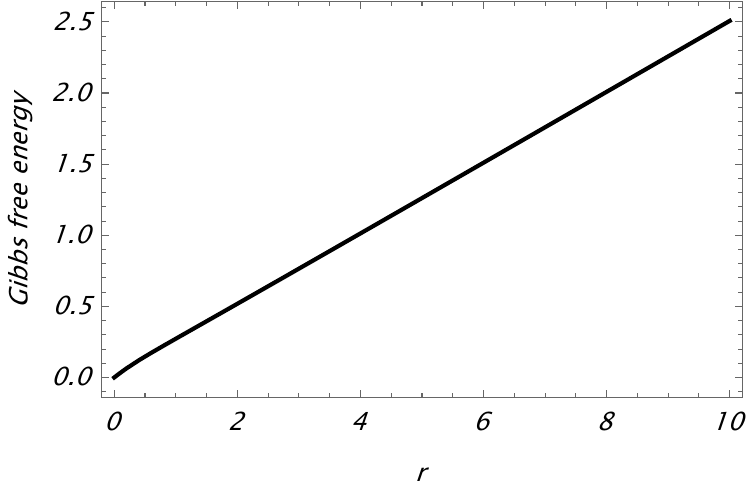}
        \caption{\textbf{Left panel}: Temperature. \textbf{Middle}: Entropy. \textbf{Right panel}: Gibbs free energy. For these plots we consider $M=1$, $b=3$, $c=-2$, $\alpha=0.2$ and $l=1$.}  
    \label{termo}
\end{figure*}

Thus, the temperature is universally suppressed relative to the Schwarzschild value 
$T_{0}=1/(4\pi r_{H})$.
This suppression is not a small correction but a structural consequence of the radial sector. 
The small expansion of $\alpha\to0$ of \eqref{eq58} yields
\begin{equation}
T_{H}
=T_{0}\left[1+\frac{(br_{H})^{c}}{2^{c+1}}l\alpha\right]+\mathcal{O}({\alpha}^{2}),
\end{equation}
showing explicitly the monotonic decrease of $T_{H}$ when considering $l\alpha<0$, see left panel in Fig. \ref{termo}.

The entropy follows from 
\begin{equation}
    dS=\frac{d{M}}{T_{H}}.
\end{equation}
Using \eqref{eq58}, one obtains after integration
\begin{equation}
\begin{split}
S=\pi r^{2}_{H}\sqrt{\frac{2^{c}+\left(br\right)^{c}l\alpha}{2^{c}}}\, {}_{2}F_{1} \left(\frac{1}{2}, \frac{2}{c}, \frac{2+c}{c},-\frac{(b r_{H})^c}{2^{c}} l\alpha\right).
\label{eq:S_final}
\end{split}
\end{equation}
Hence, the entropy is modified by the effects of MGD. 
The increase in $S$ reflects the fact that the
deformation effectively enlarges the 
horizon area by a multiplicative renormalization of the radial measure. {In the representative branch, $\{b,c\}=\{3,-2\}$, the entropy takes the following form
\begin{equation}\label{entropybranch}
    S=\frac{4\pi}{9}\left[\frac{3}{2}r_{H}\sqrt{\frac{9}{4}r^{2}_{H}+l\alpha}-l\alpha\,\text{arctanh}\left(\frac{3r_{H}}{\sqrt{\frac{9}{4}r^{2}_{H}+l\alpha}}\right)\right].
\end{equation}}
{Expanding \eqref{entropybranch} for small $\alpha$ produces
\begin{equation}
S
=S_{0}+\frac{9\pi}{2}l\alpha\left[1+\ln\left(\frac{l\alpha}{9r^{2}_{H}}\right)\right]+\mathcal{O}({\alpha}^2),
\end{equation}}
with $S_{0}=\pi r_{H}^{2}$ the classical Bekenstein-Hawking entropy. The trend of the entropy is depicted in the middle panel in Fig. \ref{termo}.

Now, the heat capacity follows the standard definition 

\begin{equation}
C_{v}=T_H\left(\frac{\partial S}{\partial T_H}\right)=T_H \frac{\left(\frac{\partial S}{\partial r_{H}}\right)}{\left(\frac{\partial T_H}{\partial r_{H}}\right)}.
\end{equation}
 
Carrying out the computation with \eqref{eq58} and \eqref{eq:S_final} leads to the exact result
\begin{equation}
C_{v}
=-\frac{12 \pi r^{3}_{H} \left[2^{c}+(b r_{H})^{c} l\alpha\right]}{2^{1+c}+(c-2)(b r_{H})^{c} l\alpha}\sqrt{\frac{1}{9r^2+4l\alpha}}.
\label{eq:C_final}
\end{equation}
This quantity remains \emph{negative}, which confirms that the BH is 
thermodynamically unstable.

The Gibbs free energy provides additional thermodynamic information. However, we first need to compute the ADM mass for this model before
obtaining the free energy. So, the ADM mass is defined by \cite{Wald:1984rg}
\begin{equation}\label{eq65}
M_{\text{ADM}}=\frac{1}{16 \pi} \lim _{r \rightarrow \infty} \int \sum_{i, j=1}^3\left(g_{i j, i}-g_{i i, j}\right) n^j d A
\end{equation}

where the integral is over a sphere of constant radius
\begin{equation}
r=\sqrt{\left(x^1\right)^2+\left(x^2\right)^2+\left(x^3\right)^2}, \quad n^j=x^j / r,
\end{equation}
and $dA$ the surface element. Here, the spatial component of the metric for large values of $r$ is given by
\begin{equation}
    g_{ij}=\left[1+\frac{2 M}{r}+\mathcal{O}\left(\frac{1}{r^2}\right)\right]\delta_{ij}.
\end{equation}
Therefore,
\begin{equation}\label{eq68}
    g_{i j, i}-g_{i i, j}=-2\partial_{j}\left[\frac{2 M}{r}+\mathcal{O}\left(\frac{1}{r^2}\right)\right].
\end{equation}
So, putting together \eqref{eq65} and \eqref{eq68} one obtains
\begin{equation}
    M_{\text{ADM}}=M.
\end{equation}
This is a non-trivial result, because some non-Schwarzschild BH such as the one reported by \cite{Alonso-Bardaji:2021yls} in the framework of quantum gravity theory, has an ADM mass which is not the parameter $M$ in the metric \cite{Balart:2024rts}.

Now, the Gibbs free energy is given by
\begin{equation}
    G=M_{\rm ADM}-T_{H}S.
\end{equation}
For the present case, $G$ has the following expression
\begin{equation}
\begin{split}
G
&=\frac{r_{H}}{4}\bigg[2-\left(1+\frac{(br_{H})^{c}l\alpha}{2^c}\right) \\& \times{}_{2}F_{1} \left(\frac{1}{2}, \frac{2}{c}, \frac{2+c}{c},-\frac{(b r_{H})^c}{2^{c}} l\alpha\right)\bigg],
\end{split}
\end{equation}
which differs markedly from the Schwarzschild expression. {Now, on the representative branch, Gibbs free energy takes the following form
\begin{equation}
\begin{split}
    G&=\frac{r_{H}}{2}-\frac{1}{108r^{2}_{H}}\bigg[27r^{3}_{H}+12r_{H}l\alpha \\&-4l\alpha\sqrt{9r^{2}_{H}+4l\alpha}\,\text{arctanh}\left(3r_{H}/\sqrt{9r^{2}_{H}+4l\alpha}\right)\bigg].
    \end{split}
\end{equation}}Its trend is shown by the right panel in Fig. \ref{termo}, corroborating the thermodynamical instability of the model.

Next, expanding for a small $\alpha$ leads to 
\begin{equation}
G=G_{0}-\frac{\left[2+\ln\left(\frac{l\alpha}{9r^{2}_{H}}\right)\right]l\alpha}{18r_{H}}+\mathcal{O}({\alpha}^{2}),
\end{equation}
with $G_{0}\equiv r_{H}/4$.

The thermodynamic portrait that emerges is therefore unambiguous, the parameter 
$l\alpha$ acts coherently across all quantities by lowering the temperature, increasing 
the magnitude of entropy and heat capacity and modifying the Gibbs energy in a 
qualitatively new way that cannot be reproduced by a simple rescaling of the Schwarzschild 
mass.  
Every correction vanishes smoothly in the limit $\alpha\to0$, but for finite 
$\alpha$ and $l$ the deviations are systematic.

\section{Conclusions}\label{section6}

In this work, we have constructed and analyzed a new family of static and
spherically symmetric geometries generated through the MGD approach and supported by a nonlinear scalar-vector sector.
Starting from the Schwarzschild solution as the seed geometry and imposing
a general algebraic relation among the components of the decoupling tensor,
we obtained an exact analytical deformation of the radial metric component,
leading to a non-Schwarzschild spacetime while preserving the Schwarzschild
temporal potential. This provides,
to our knowledge, the first exact non-Schwarzschild BH sourced by a genuine
scalar-vector system within the framework of GD.

A central result of this work is that the resulting solution space rather than describing a unique BH
configuration, the MGD deformation naturally gives rise to two distinct
BH branches, denoted throughout this work as B1
($l\alpha>0$) and B2 ($l\alpha<0$). Although both branches preserve the
same Schwarzschild event horizon at $r_H=2M$ and remain asymptotically
flat, they exhibit remarkably different physical properties. The B2 branch
satisfies both the odd- and even-parity stability criteria but develops an
effective matter sector violating the weak energy condition in the exterior
region. Conversely, the B1 branch admits a real scalar-field
reconstruction together with a more satisfactory effective energy-momentum
tensor, although it does not satisfy the odd-parity stability criterion.
Therefore, neither branch is universally preferred, revealing a nontrivial
interplay between dynamical stability, energy conditions and causal structure. 

The causal analysis also shows that the parameter space contains more than
one geometrical realization. Within the parameter domain considered in this
work, both B1 and B2 describe genuine BH geometries characterized
by a single Killing horizon and the absence of additional throats or
signature changes. Outside the BH parameter domain of B2, however, the
same analytical family admits wormhole configurations, illustrating the
rich geometrical structure generated by the MGD deformation.

An important outcome of the present analysis is the explicit reconstruction
of the matter sector supporting the B1 branch. We showed that the scalar
field can be obtained analytically in terms of incomplete elliptic
functions, while the nonlinear electromagnetic invariant, the interaction
potential and the electric potential are completely reconstructed from the
field equations without introducing additional ansatz. The scalar field
remains regular throughout the exterior region and approaches a constant at
spatial infinity, whereas the nonlinear electromagnetic sector decays
rapidly, preserving the Schwarzschild asymptotics of the geometry. These
results demonstrate that the anisotropic source generated by GD admits a consistent interpretation as a genuine nonlinear
scalar-vector system rather than as an effective anisotropic fluid.

The geodesic analysis further shows that, although the temporal component
of the metric remains unchanged, the radial deformation modifies the
relation between radial and angular motion, producing observable deviations
from Schwarzschild in bounded, scattering and plunge trajectories. Hence,
the MGD deformation leaves the qualitative orbital structure unchanged
while generating measurable strong-field corrections associated with the
scalar-vector interaction.

The thermodynamic properties of the solution were also investigated in
detail. The Hawking temperature, entropy, heat capacity and Gibbs free
energy were obtained analytically, showing that the deformation introduces
nontrivial corrections while preserving the ADM mass of the spacetime. This
behavior reflects the fact that the scalar and vector fields do not
generate an independent asymptotic gravitational charge but instead
produce a purely radial deformation of the geometry. In this sense, the
matter sector exhibits a stealth-like behavior, remaining gravitationally
active in the strong-field region while becoming asymptotically invisible.

More generally, our results suggest that GD may naturally generate rich families of exact solutions whose physical interpretation cannot be inferred from the geometry alone, but instead requires the combined analysis of causal structure, perturbative stability and matter reconstruction. The coexistence of physically distinct BH
branches, together with the possibility of wormhole geometries outside the
BH domain, reveals that MGD deformations encode a much broader
phenomenology than previously recognized. These results open new
perspectives for investigating the relation between energy conditions,
perturbative stability, nonlinear matter couplings and strong-field
gravitational physics within exact analytical models.

\section*{ACKNOWLEDGMENTS}

Y. Gómez-Leyton acknowledges to ANID subvención en la academia convocatoria año 2025 85250186.

\bibliography{biblio.bib}
\bibliographystyle{elsarticle-num}

\end{document}